\def\utwothirtyeight{$\rm ^{238}U$~}
\def\thtwothirtytwo{$\rm ^{232}Th$~}
\def\cosixty{$\rm ^{60}Co$~}

\documentclass{JINST}
\title{Study of fast neutron detector for COSINE-100 experiment}

\author{COSINE-100 Collaboration}
\author{G.~Adhikari$^a$,  P.~Adhikari$^a$, C.~Ha$^{b}$, E.J.~Jeon$^{b}$, K.W.~Kim$^{b}$,
  N.Y.~Kim$^{b}$, Y.D.~Kim$^{a,b}$, Y.J.~Ko$^{b}$, H.S.~Lee$^{b}$, J.~Lee$^{b}$, M.H~Lee$^{b}$, R.H.~Maruyama$^{c}$,
  S.L.~Olsen$^{b}$, and H.~Prihtiadi$^{b,d}$ \\
 \llap{$^a$} Department of Physics, Sejong University, Seoul 05006, Korea\\
  \llap{$^b$} Center for Underground Physics, Institute for Basic Science~(IBS), Daejeon 34047, Korea\\  
  \llap{$^c$} Department of Physics, Yale University, New Haven, CT 06520, USA\\
  \llap{$^d$} Department of Physics, Bandung Institute of Technology, Bandung 40132, Indonesia\\
 
E-mail: \email{yjko@ibs.re.kr}}

\abstract{
  A monitoring system for fast neutrons is planned in the COSINE
  experiment, a dark matter experiment with NaI crystals. We pursued
  several R\&D approaches for a neutron detector using a liquid
  scintillator (LS). A pulse shape discrimination (PSD) technique is
  used for the identification of neutron events and the PSD properties
  of two different LS were compared. A good separation power between
  neutrons and $\gamma$ has been achieved for energies between
  200~keVee to 1500~keVee. The combination of alumina adsorption,
  filtration, and water extraction is effective in purifying the LS,
  which leads to a reduction in the $\alpha$ contamination by
  $^{210}$Po of more than a factor of two. The measured activities of
  the internal $\alpha$ are 0.36$\pm$0.04~mBq/kg and
  0.21$\pm$0.03~mBq/kg before and after purification, respectively.
}

\keywords{Neutrons; COSINE-100 Experiment; Pulse shape discrimination;  Dark Matter; Liquid Scintillator; NMD}

\begin{document}
\section{Introduction}
\label{sec:intro}

The DAMA/LIBRA experiment, which has been operational for over 15
years, is an experiment to search for the evidence of dark matter
scattering off nuclei using an array of ultra-low-background NaI(Tl)
crystals~\cite{DAMA_01}. DAMA/LIBRA consistently reports a positive
signal with a significance of 9.3$\sigma$ for an annual
modulation~\cite{DAMA_02,DAMA_03}, and the signal is consistent with a
weakly interacting massive particle (WIMP) that has a spin-independent
nucleon cross section of approximately
2$\times$10$^{-40}$~cm$^{2}$~\cite{savage}.

The COSINE-100 experiment~\cite{cosine3} has developed an
ultra-low-background NaI (Tl) crystal detector~\cite{
  kimsnai1,cosine1,cosine2} to search for WIMPs. The primary goal of
the experiment is to confirm or refute the claimed observation of the
DAMA/LIBRA experiment with the same type of crystal. The COSINE-100
experiment is located in the Yangyang Underground Laboratory (Y2L)
with a minimum earth overburden of about 700 m~\cite{cosine4}. There
have been continuing debates about whether or not the DAMA/LIBRA
annual modulation signal is due to muon-induced neutron signals that
may have been  seasonally modulated ~\cite{Blum_Kfir}. Therefore, it
is important to understand the environmental neutron background for
the COSINE-100 experiment.

In the case of fast neutrons, the energy conversion process is elastic
scattering with light nuclei that gives rise to recoil nuclei. The
main sources of neutrons are the spontaneous fission of
\utwothirtyeight in the rocks, ($\alpha$, n) reactions caused by
$\alpha$ particles from the decay of \utwothirtyeight and
\thtwothirtytwo, and cosmic ray muons that split the target
nuclei~\cite{kims_csi2}. Highly energetic muons interacting with the
shielding materials can generate neutrons inside the shield. Since
modulation of the environmental neutrons that can interact with nuclei
is possible, neutron monitoring is necessary for the annual modulation
study. In order to do that, we are constructing a neutron monitoring
detector (NMD) using a liquid scintillator (LS), and it will be
installed inside the main shielding of the COSINE detector. Here, we
describe the R\&D in the construction of the NMD for the COSINE-100
experiment.

\section{Neutron Detector and Experimental Setup}
\label{sec:detector}

A cylindrical vessel of 5-cm length and 4.5-cm inner diameter made of
1.5-cm thick Teflon is used for the neutron detector. An acrylic
window of 10-mm thickness is attached to each end of the vessel with
suitable O-rings to prevent leakage. Two different LSs are the
candidates for the target of the neutron detector. One is based on
Linear Alkyl Benzene (LAB) and the other is a commercial product
called ``Ultima Gold-F (UG-F)'' based on Di-isopropylnaphthalene (DIN;
C$_{16}$H$_{20}$). The test detector is read out by two 3-inch
photomultiplier tubes (PMT; Hamamatsu R12669SEL) mounted on each end
of the cylinder. The PMTs have bi-alkali photocathodes that have 35\%
maximum quantum efficiency at a wavelength of around 420~nm.

\begin{figure*}[b]
  \begin{center}
    \includegraphics[width=.7\columnwidth]{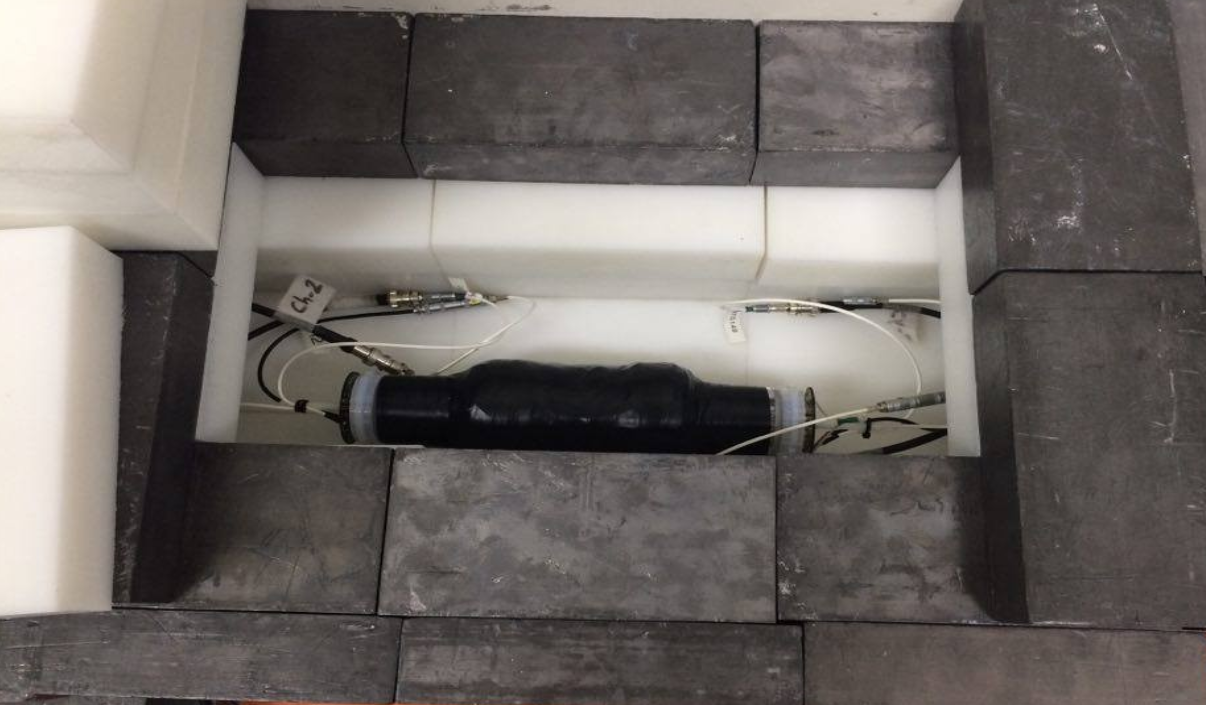}
    \caption{Experimental setup for the PSD of the neutron detector.}
    \label{fig:exp_setup}
  \end{center}
\end{figure*}

The flash ADC, which has a 12-bit resolution with 500~MHz sampling,
records the digitized signal waveforms from the anode readout of two
PMTs. We use the logic "AND" of the signals from the two PMTs to form
a trigger signal. Only when both PMTs simultaneously have signals
above the threshold are the waveforms are recorded. The trigger
threshold is set at 9.2~mV when two channels coincide within a 64~ns
window. The SY4252 CAEN module is used to supply the high voltage on
PMTs. In order to reduce environmental neutrons and $\gamma$, the
detector was surrounded by 5-cm thick polyethylene (PE) and a 5-cm
thick normal lead shield in the test setup, as shown in
Fig. \ref{fig:exp_setup}.

There is another similarly configured detector with a ten times larger
volume to understand the background level. The detector is installed
inside the KIMS-CsI shielding test facility at the
Y2L~\cite{kims_csi}. The available polyethylene shielding may be good
enough to prevent external neutrons so that neutron-like events in the
$\alpha$ band are assumed to be negligible.

\subsection{Liquid Scintillator}
\label{sec:LS}

We chose LAB as the base solvent material. LAB has several advantages
as follows:
\begin{itemize}
\item Light yield comparable with pseudocumene.
\item High flash point (approximately 130$^\circ$C).
\item Long attenuation length (> 10~m at 430-nm wavelength)~\cite{Ding_Y}.
\item Domestically available.
\item Relatively safe material.
\end{itemize}
The emission spectrum of LAB has a maximum at 340~nm, so we mix the
solvent with the wavelength shifter to adjust the wavelength of the
optical photons to be suitable for the PMT. We used the neutrino grade
of the 2,5 diphenyloxazole (PPO; C$_{15}$H$_{11}$NO) for the primary
fluor and the scintillation grade of the
1,4-bis(2-methylstyryl)benzene (bis-MSB;
(CH$_3$C$_6$H$_4$CH=CH$_2$)$_2$C$_6$H$_4$) for the secondary
wavelength shifter. The amount of PPO was chosen to be 3~g/L and of
bis-MSB was chosen to be 30~mg/L~\cite{JS_Park_1}.

The commercial LS cocktail produced by PerkinElmer company known as
``Ultima Gold-F,'' which has a higher flash point (approximately
150$^\circ$C), is also studied. DIN is a solvent for this cocktail. In
comparison to cocktails based on DIN, the classical cocktails show
very short pulse lengths~\cite{Song_H}. The discrimination is much
easier if the pulses are extended. As UG-F is a commercial LS, the
type and quantity of the wavelength shifter mixed into the DIN solvent
were not clear.

\section{Pulse Shape Discrimination} \label{sec:PSD}

\subsection{Energy Calibration of the Detector} \label{sec:calibration}

\begin{figure*}[!t]
  \begin{center}
    \includegraphics[width=.7\columnwidth]{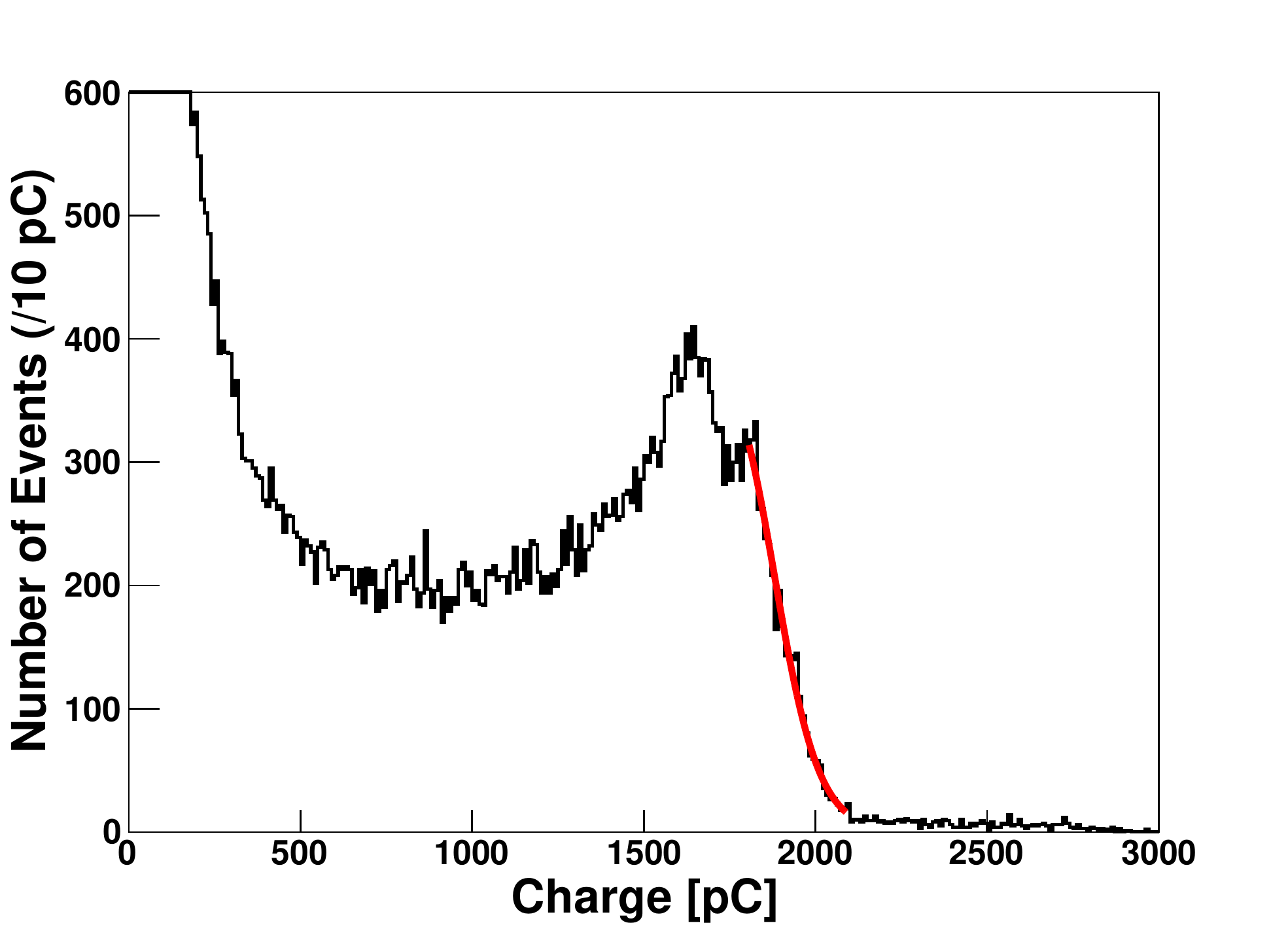}
    \caption{Charge distribution of $\gamma$ events from \cosixty for energy
      calibration. The red line is a fitted function to estimate the energy 
      of the Compton edge of 1332~keV $\gamma$ from \cosixty. Both
      Compton edges of 1332~keV and 1172~keV $\gamma$ are
      superimposed as a single Compton edge.}
    \label{fig:calibration_co60}
  \end{center}
\end{figure*}

The detector was calibrated with 1332-keV $\gamma$ from
\cosixty. Since it is difficult to find the full peak of the \cosixty
charge distribution due to the small size of the detector, the Compton
edge, 1122~keV for 1332 keV $\gamma$, was used for the energy
calibration. The following function is used for fitting the charge
distribution to estimate the energy of the Compton edge:
\begin{equation}
  \label{eq:comptonedge}
  f(q)=\frac{p_0}{\exp\left[p_1(q-p_2)\right]+1},
\end{equation}
where the $p_i$s are the fitting parameters and $q$ is the charge. In
this equation, $p_2$ denotes the energy of the Compton
edge. Fig. \ref{fig:calibration_co60} shows the charge distribution of
the $\gamma$ events from \cosixty and the fitted function.

\subsection{PSD Parameter and Optimization}

When a particle deposits its energy in the LS, it excites the solvent
molecules, which causes chemical quenching. The energy of the solvent
molecules is transferred to the flour molecules, and optical photons
emitted from the flour molecules are received by the photocathode of
the PMT. The light emitted from the vast majority of the organic LS
consists of two main components: the fast and the slow components. The
fast one has an exponentially decaying lifetime that is typically in
the range of a few nanoseconds. The slow component has an exponential
tail extending out to several hundred nanoseconds~\cite{craun_R}. The
relative intensities of the two components depend on the specific
energy loss of the particle passing through the LS.

\begin{figure*}[!b]
  \begin{center}
     \begin{tabular}{cc}
  \includegraphics[width=.4\columnwidth]{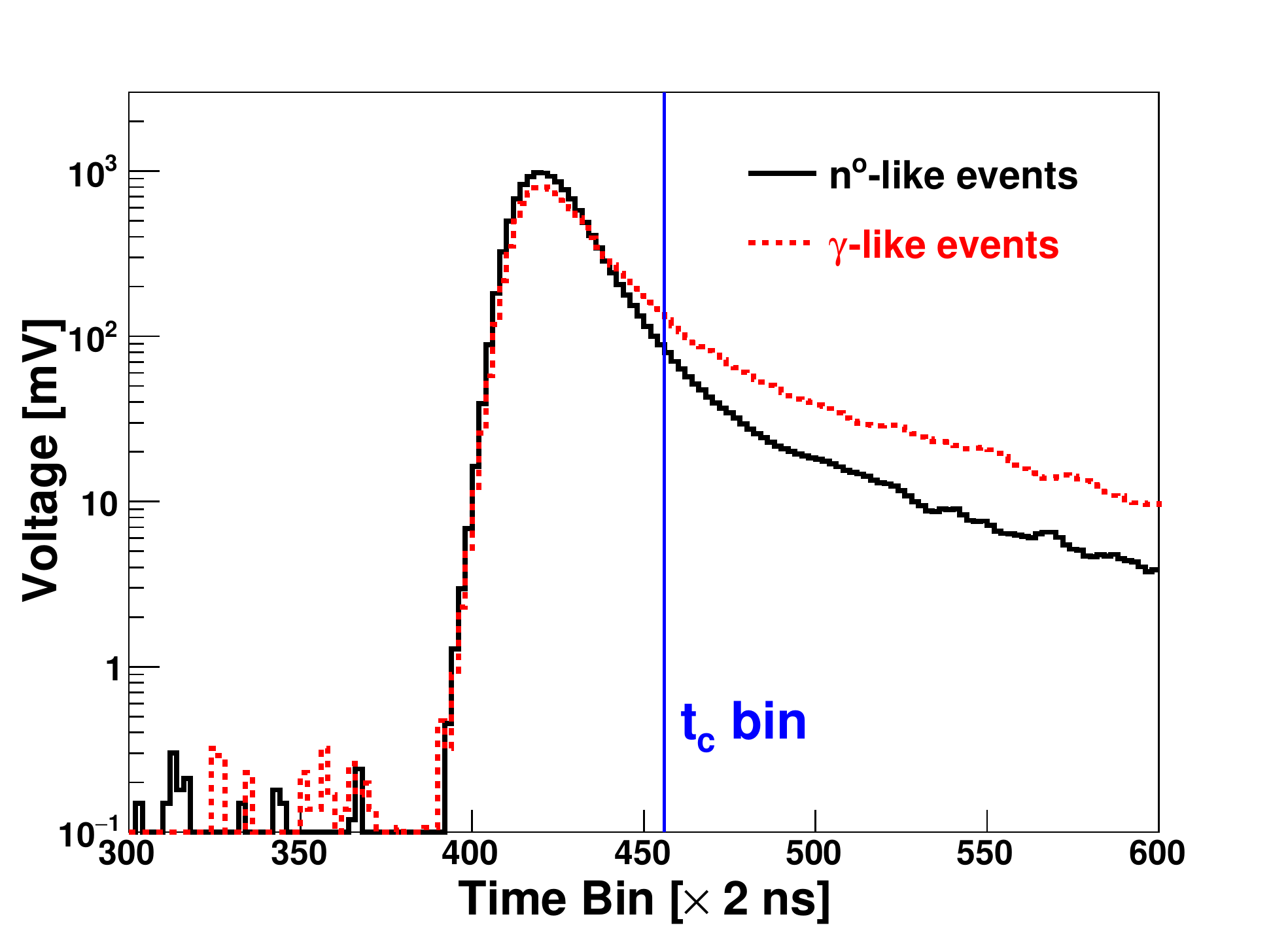} &
      \includegraphics[width=.4\columnwidth]{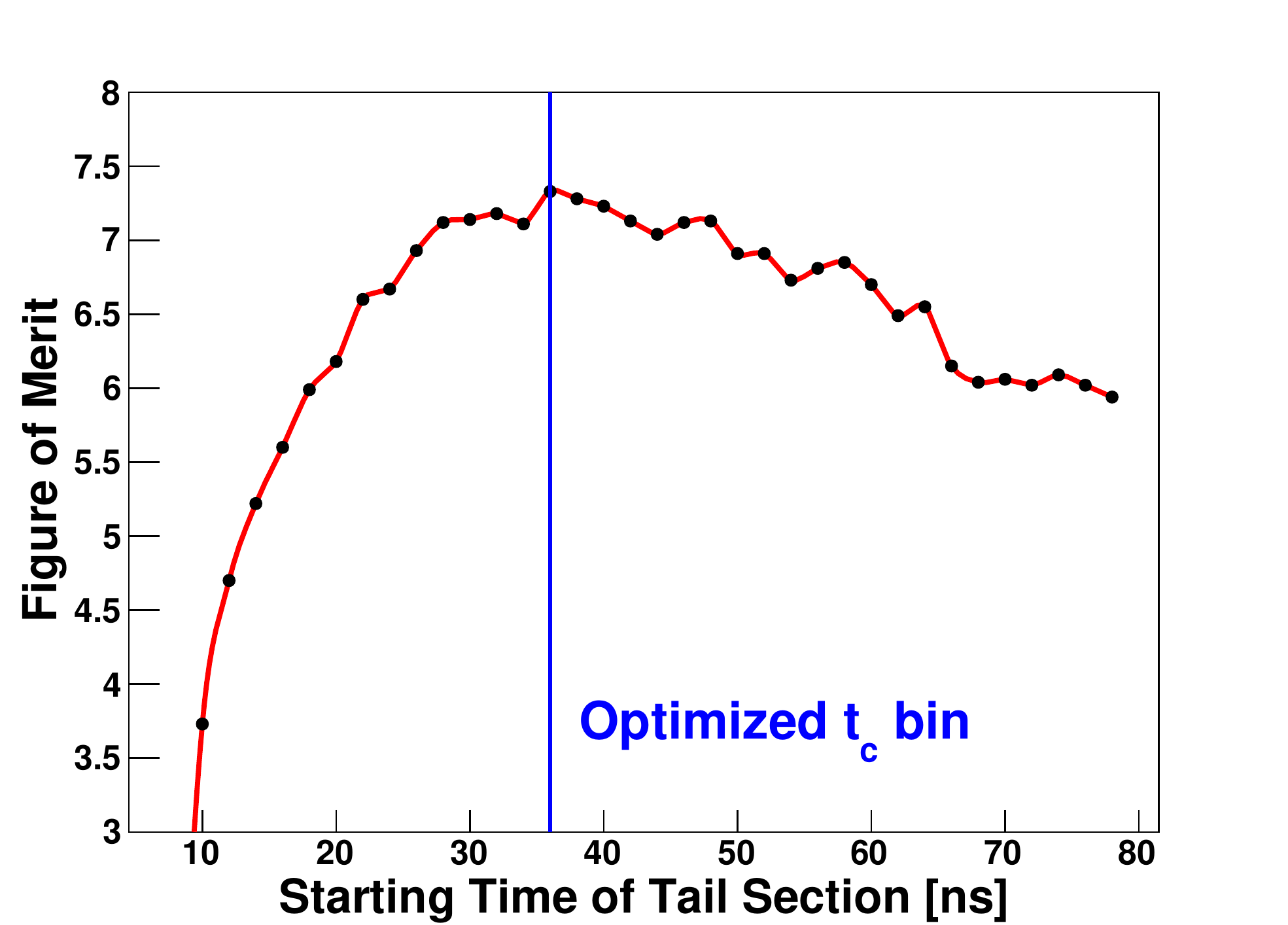} \\
      (a) & (b) \\
   \end{tabular}
 \caption{(a) Signal shape of neutron and $\gamma$ events in UG-F.
      (b) Optimization of starting time of the tail section, $t_c$.}
     \label{fig:pulse_shape}
  \end{center}
\end{figure*}

Fast neutrons deposit their energies via proton recoil while $\gamma$
deposit their energies via electron recoil. The energy losses of the
proton and the electron are different, so the signal shape of fast
neutron events is different from that of $\gamma$ events, as shown in
Fig. \ref{fig:pulse_shape}~(a). The signal shape of proton recoil has
a longer tail than that of electron recoil due to greater
de-excitation of different states in the
LS~\cite{Lombardi_P}. Therefore, the $\gamma$ and neutron events can
be distinguished via their signal shape.

For particle identification, the tail charge $Q_{\mathrm{tail}}$ is
defined as
\begin{equation}
  \label{eq:qtail}
  Q_{\mathrm{tail}}=\int_{t_c}^{t_{\mathrm{max}}}\frac{V(t)}{R_{\mathrm{terminal}}}dt,
\end{equation}
where $R_{\mathrm{terminal}}(=50~\Omega)$ is the terminal resistance
and the range [$t_c$, $t_{\mathrm{max}}$] is the tail section. The
$\gamma$ and neutron events can be separated using the ratio between
the tail charge $Q_{\mathrm{tail}}$ to the total charge
$Q_{\mathrm{total}}$; thus, this ratio is used as a PSD parameter.

The starting time of tail section, $t_c$ is optimized to get highest
separation power between the $\gamma$ and neutron bands via neutron
calibration data. The separation power is quantitatively expressed in
terms of the figure of merit (FoM) as follows:
\begin{equation}
  \label{fomcalculation}
  \mathrm{FoM}=\frac{\left|m_n-m_\gamma\right|}{\sqrt{\sigma_n^2 + \sigma_\gamma^2}},
\end{equation}
where $m_n$ ($m_\gamma$) and $\sigma_n$ ($\sigma_\gamma$) represent
the mean and standard deviation of the PSD distribution of the
neutrons ($\gamma$s), respectively. The higher FoM indicates better
discrimination between events induced by $\gamma$ and
neutron. Therefore, we select the $t_c$ to be 36 ns because the FoM
has the maximum value at $t_c = 36$~ns from the maximum pulse bin, as
shown in Fig. \ref{fig:pulse_shape}~(b).

\subsection{PSD Performance of Different Liquid Scintillators}

\begin{figure*}[!b]
  \begin{center}
    \begin{tabular}{cc}
      \includegraphics[width=.4\columnwidth]{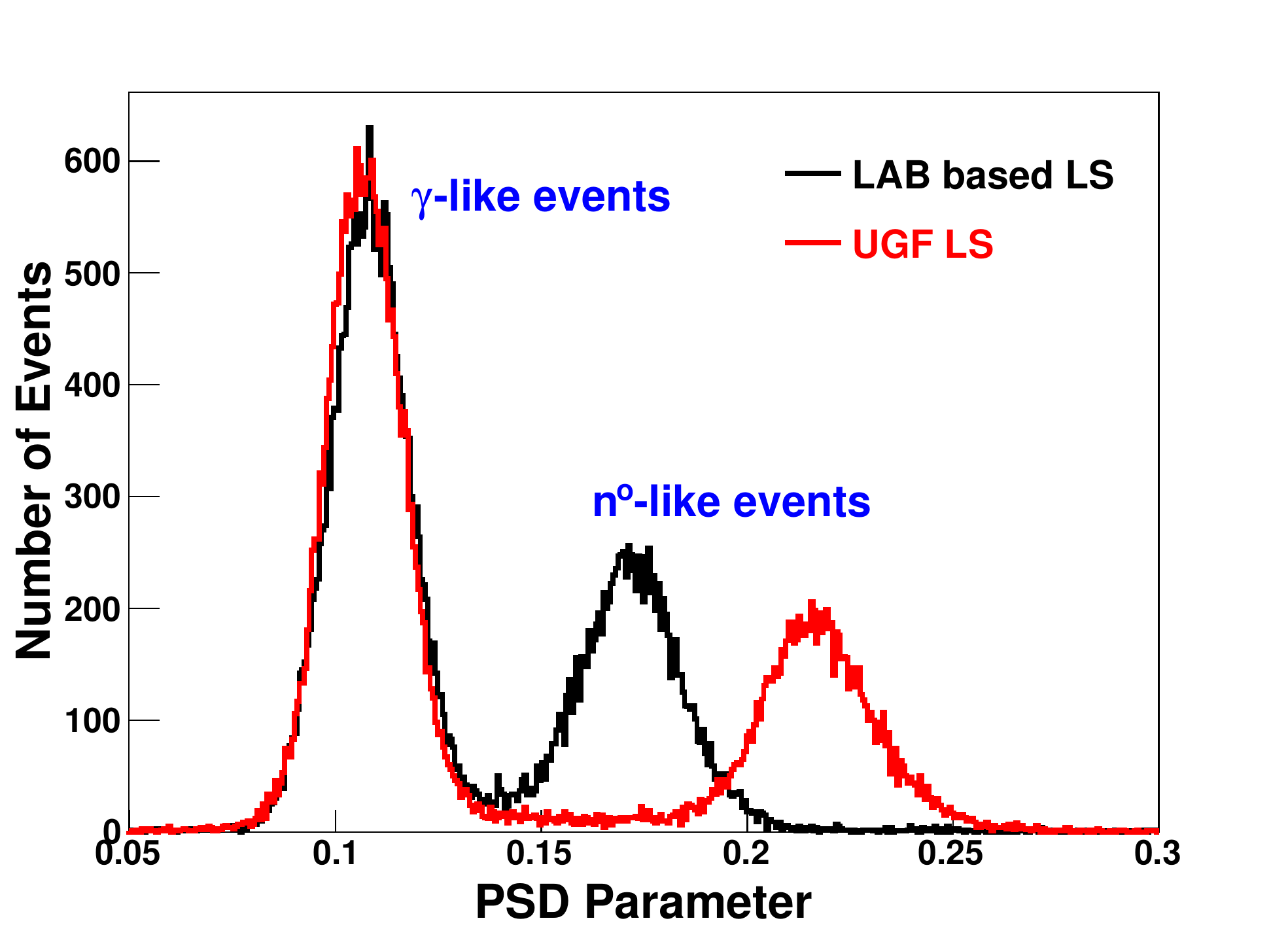} &
      \includegraphics[width=.4\columnwidth]{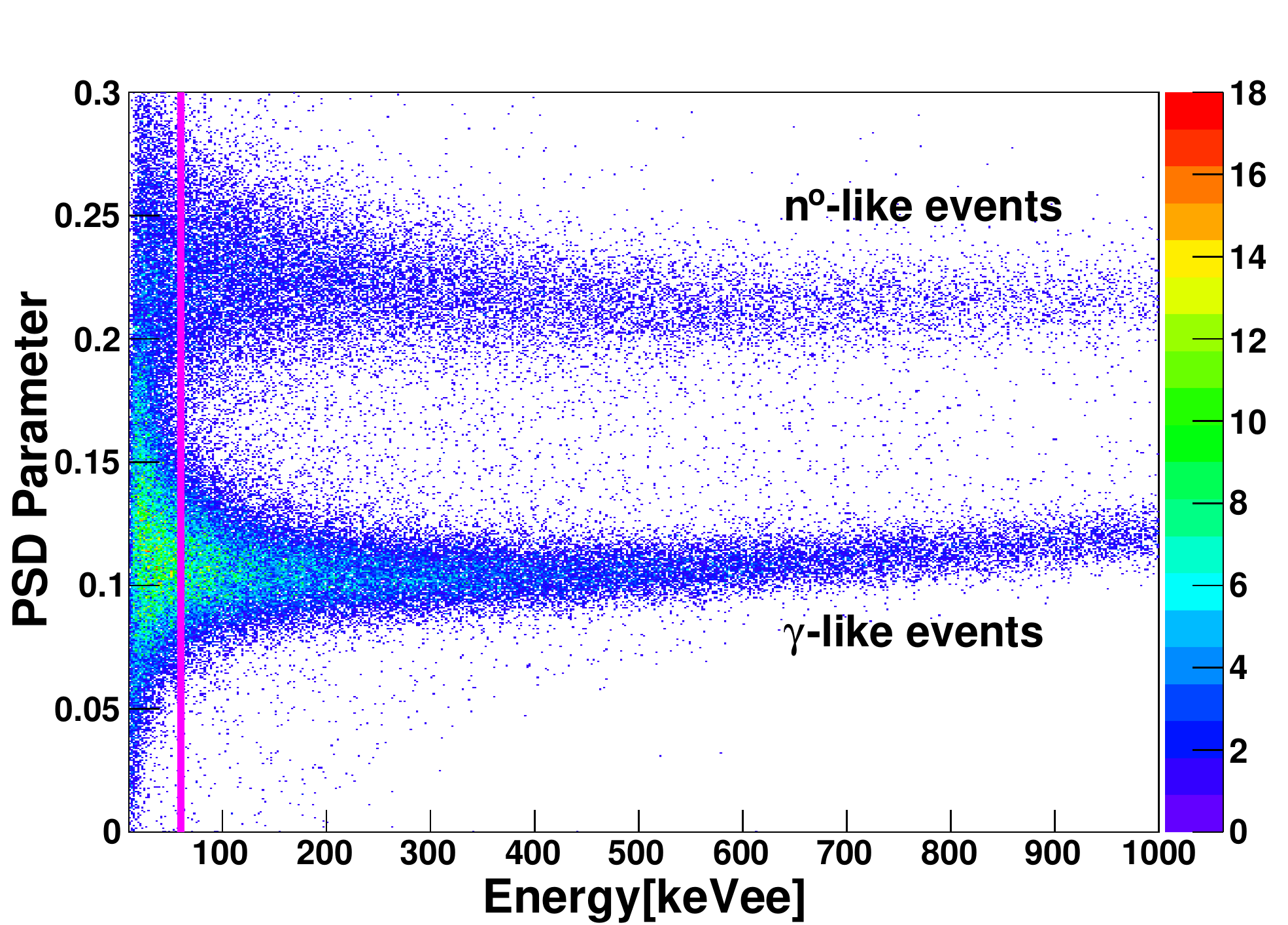} \\
         (a) & (b) \\
   \end{tabular}
     \caption{ (a) Comparison of the PSD parameter distributions of LAB-
      and DIN-based (UG-F) LSs in the energy range [200,~1000] keVee, and
      (b) scatter plot between the PSD parameter and energy in the case of the UG-F.
      The magenta line shows the PSD threshold.}
    \label{fig:psd_comparision}
  \end{center}
\end{figure*}

The LS is filled into the container as described in
Sec. \ref{sec:detector}. If there are oxygen molecules in the LS,
since they may remain between the scintillating molecules, they can
lower the light output and degrade PSD performance. In order to
control the oxygen quenching, we fill the LS into the container in an
oxygen-free environment inside the glove box. The separation power
between neutron and $\gamma$ events from a $^{252}$Cf source is
compared between the samples via the technique described in the
previous section. In this comparison, the $Q_{\mathrm{tail}}$ is
optimized for each sample and the PSD parameter works well for both
samples, as shown in Fig. \ref{fig:psd_comparision}~(a).

The FoM of UG-F is measured to be 7.1 in the energy range
[200,~1000]~keVee, while the LAB-based LS shows an FoM of 4.01
measured in the same energy range. The PSD threshold is approximately
70~keV with a 3.1~FoM for the small 70~ml R\&D detector, as shown in
Fig. \ref{fig:psd_comparision}~(b). So, we expect acceptable an FoM
and PSD threshold for the COSINE NMD having mass of about 5~kg. Based
on the PSD performance, UG-F is a better candidate material for the
NMD.

\section {Internal Alpha Background}
\label{sec:bkgd}

\begin{figure*}[b]
  \begin{center}
    \includegraphics[width=.7\columnwidth]{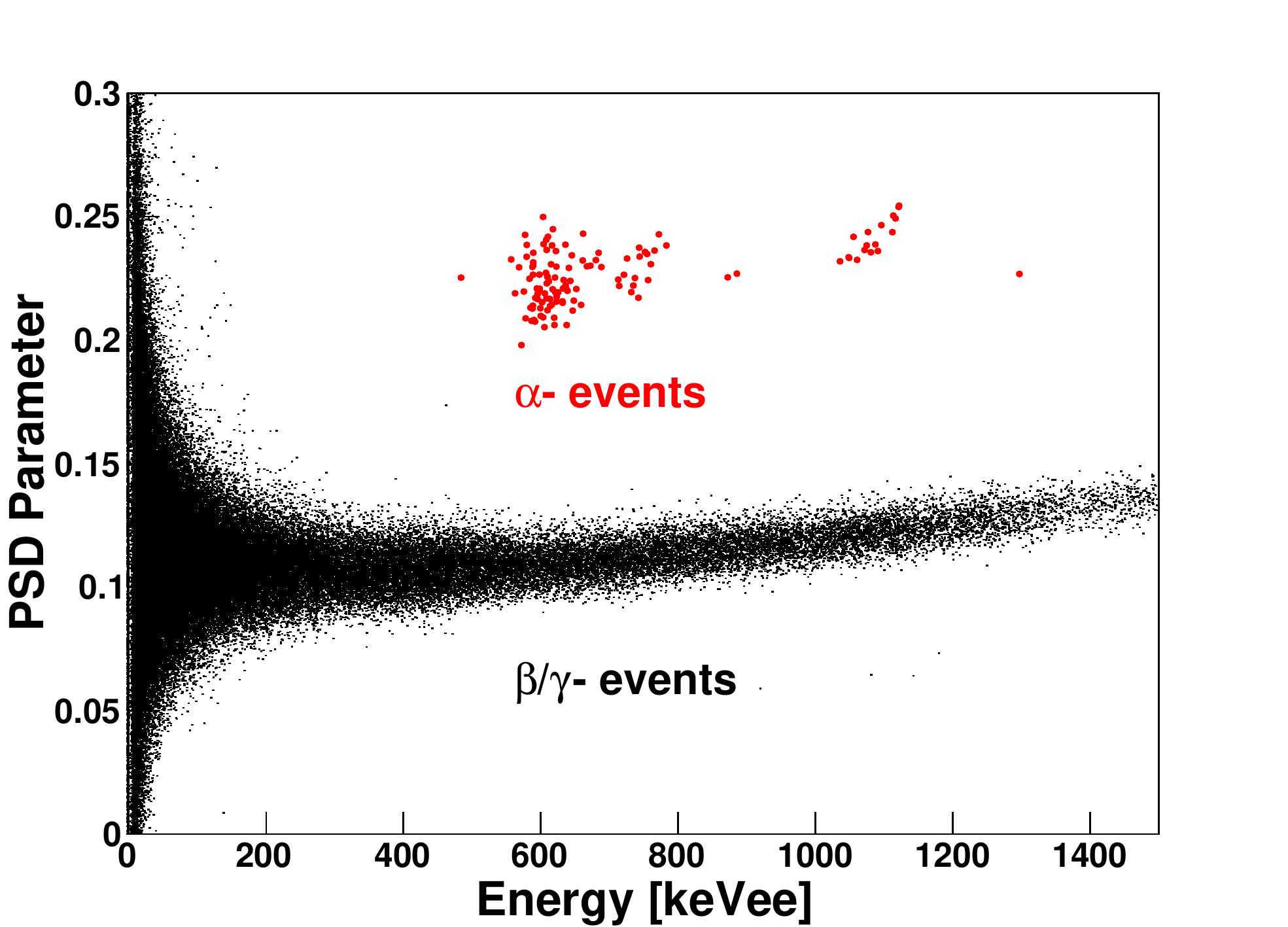} 
    \caption{Scatter plot between PSD parameter and electron
     -equivalent energy for background data}
    \label{fig:bkg_psd_data}
  \end{center}
\end{figure*}

Since the energy loss of an $\alpha$ particle is similar to that of a
proton, $\alpha$ events cannot be distinguished from neutron events
via PSD. It means that the internal $\alpha$ background will give rise
to uncertainties in the neutron rate measurement, so understanding the
$\alpha$ background and reducing it as much as possible is very
important for neutron measurements. In order to understand the
background, a detector filled with UG-F having a volume of
approximately 700~ml is installed inside the KIMS-CsI shielding test
facility at the Y2L.

\subsection{Identification of Alpha Sources}
\label{sec:id_alpha}

\begin{figure*}[!t]
  \begin{center}
    \begin{tabular}{ccc}
    \includegraphics[width=.3\columnwidth]{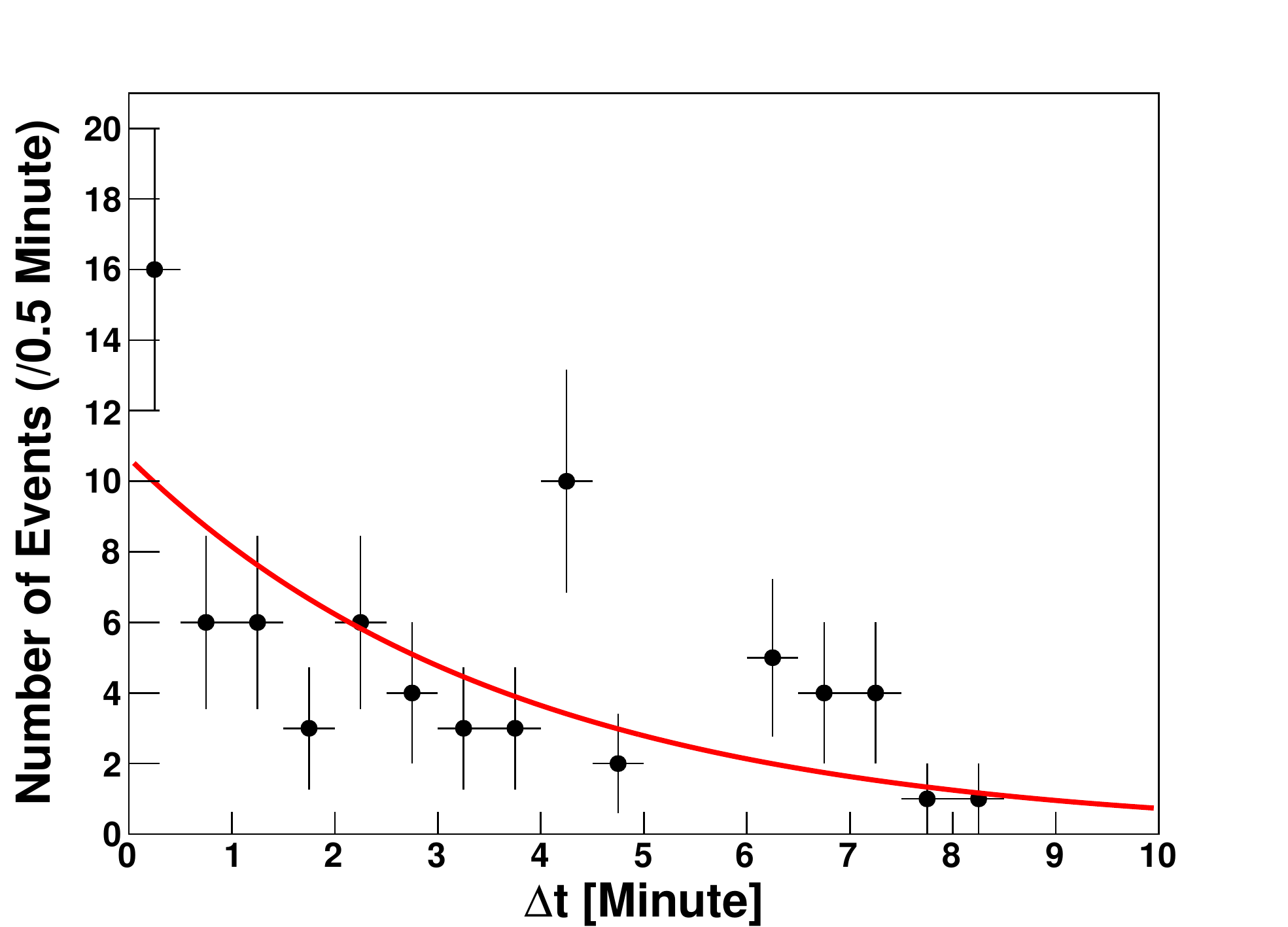}&
    \includegraphics[width=.3\columnwidth]{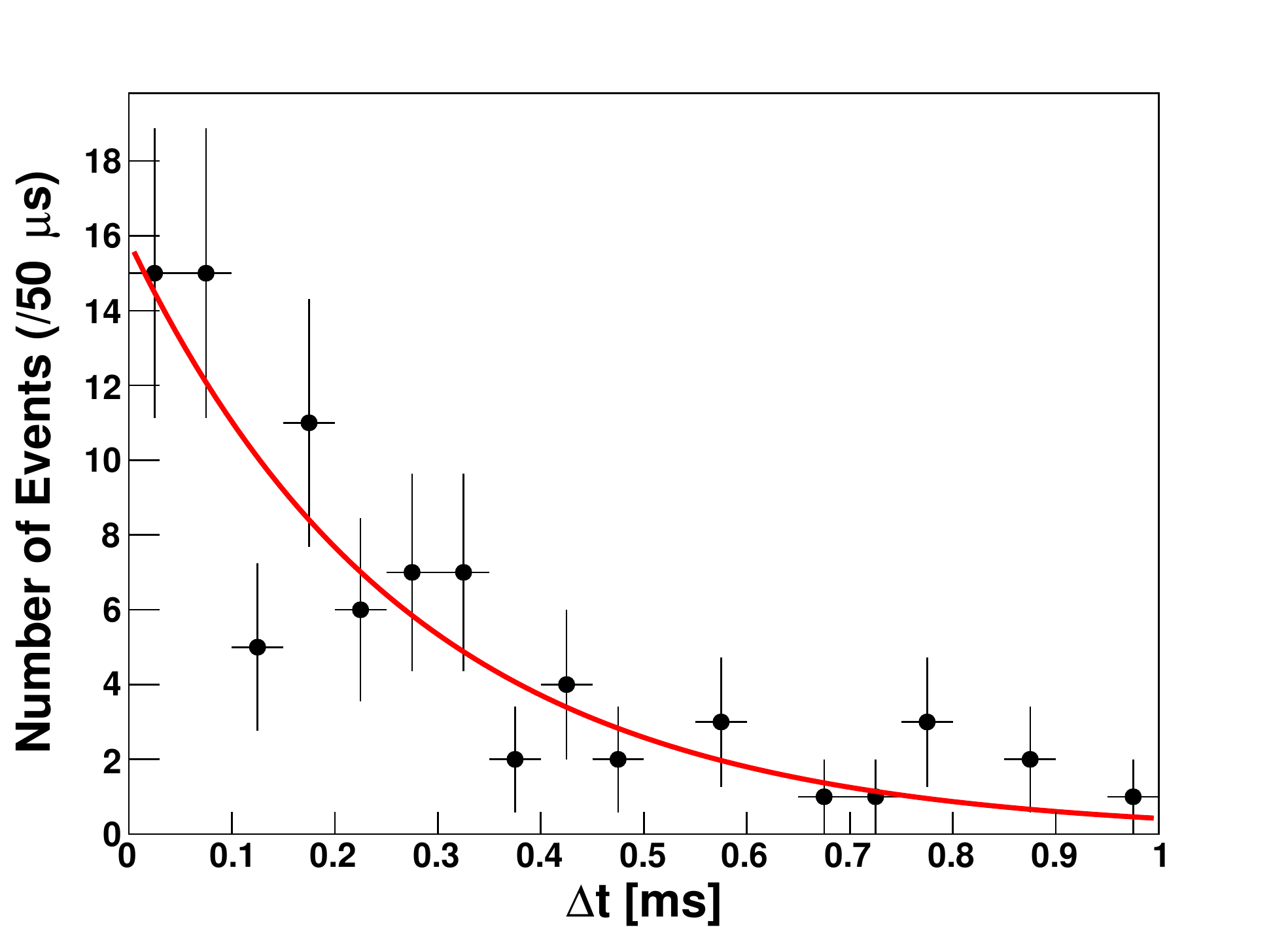}&
    \includegraphics[width=.3\columnwidth]{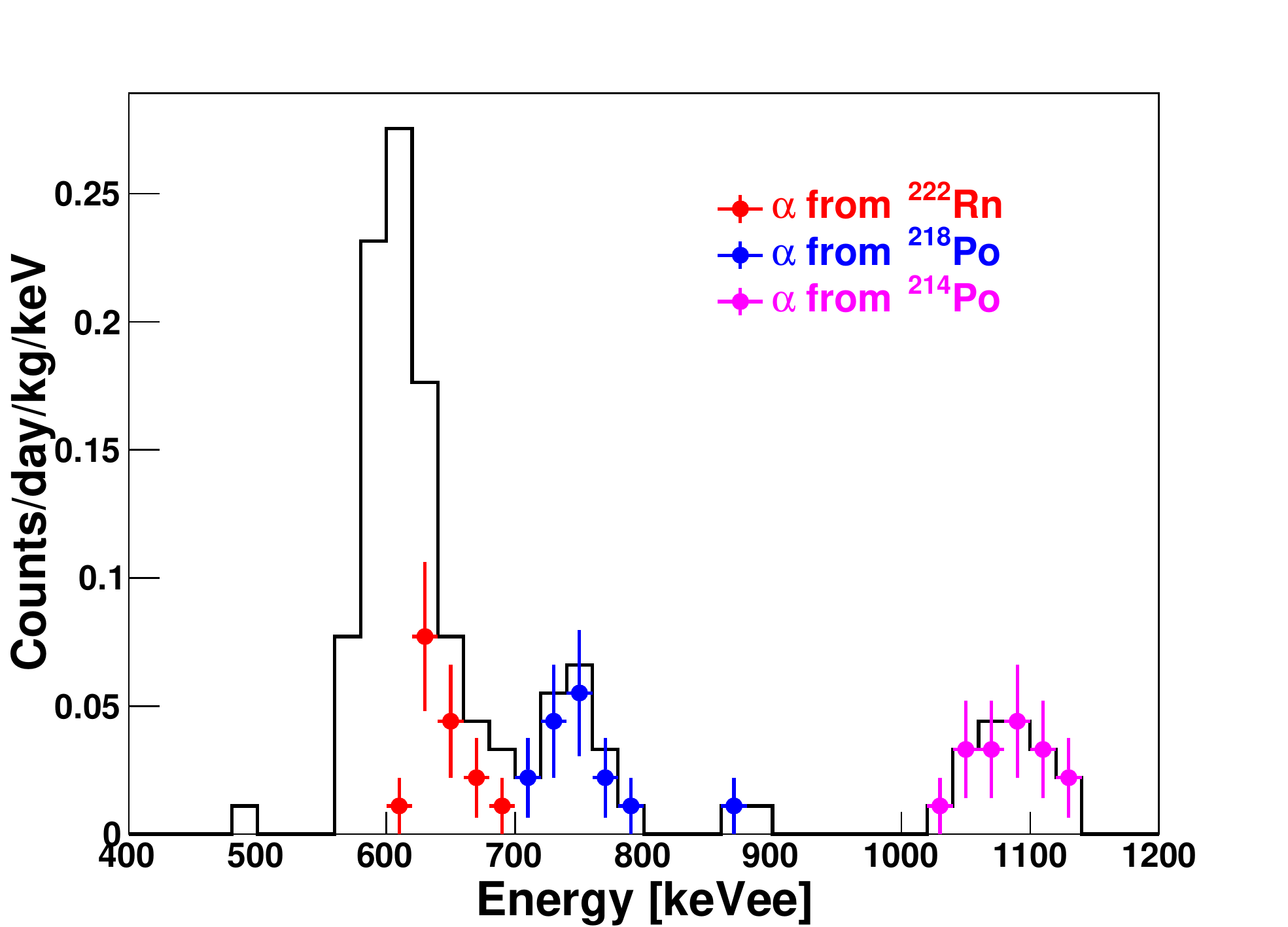}\\
         (a) & (b) & (c)\\
      \end{tabular}
      \caption{(a) Time difference distribution between two $\alpha$ events from
    $^{222}$Rn and $^{218}$Po. (b) Time differences between $\beta$ and
    $\alpha$ events from $^{214}$Bi and $^{214}$Po, respectively.
    The red curves in (a) and (b) are an exponential function fitted to the time 
    difference. (c) Energy distribution of the $\alpha$ events. }
    \label{fig:decay_time}
  \end{center}
\end{figure*}

In Sec.\ref{sec:PSD}, the DIN-based LS (UG-F) is selected for the NMD,
thus the activity of the internal $\alpha$ from the UG-F are measured
at the Y2L. Since the PSD of $\alpha$ events is similar to that of
neutron events, the $\alpha$ events can be identified via the PSD from
$\beta$/$\gamma$ events. In Fig.\ref{fig:bkg_psd_data}, one can see a
band and several  islands. The band denotes $\beta/\gamma$ events, and
the islands are $\alpha$ events.

The $\alpha$-$\alpha$ coincidence between $^{222}$Rn and $^{218}$Po
decay and $\beta$-$\alpha$ coincidence between $^{214}$Bi to
$^{214}$Po decay can be identified because of their fast
lifetimes~\cite{cosine1,kims_csi}. Fig. \ref{fig:decay_time} shows the
distribution of the measured time intervals between two successive
$\alpha$-$\alpha$ and $\beta$-$\alpha$ induced events. The lifetime
estimated from the fitting of $^{218}$Po is 2.6$\pm$0.5~min,  and that
of $^{214}$Po is obtained as 191$\pm$26~$\mu s$, which are consistent
with the known values. The red and blue dots in
Fig.\ref{fig:decay_time} (c) show the energy distributions of tagged
$\alpha$ events from $^{222}$Rn and $^{218}$Po,
respectively. Likewise, the magenta dots in Fig.\ref{fig:decay_time}
(c) show the energy distributions of tagged $\alpha$ events from
$^{214}$Po.

Similarly, we can identify an $\alpha$-$\alpha$ coincidence with a
halflife of 145~ms between $^{220}$Rn to $^{216}$Po in the $^{232}$Th
decay chain. However, the number of events after applying the
selection criteria is too small to fit, so we calculated
conservatively and obtained an upper limit of 0.01 mBq/kg.

\begin{figure*}[!b]
  \begin{center}
    \includegraphics[width=.7\columnwidth]{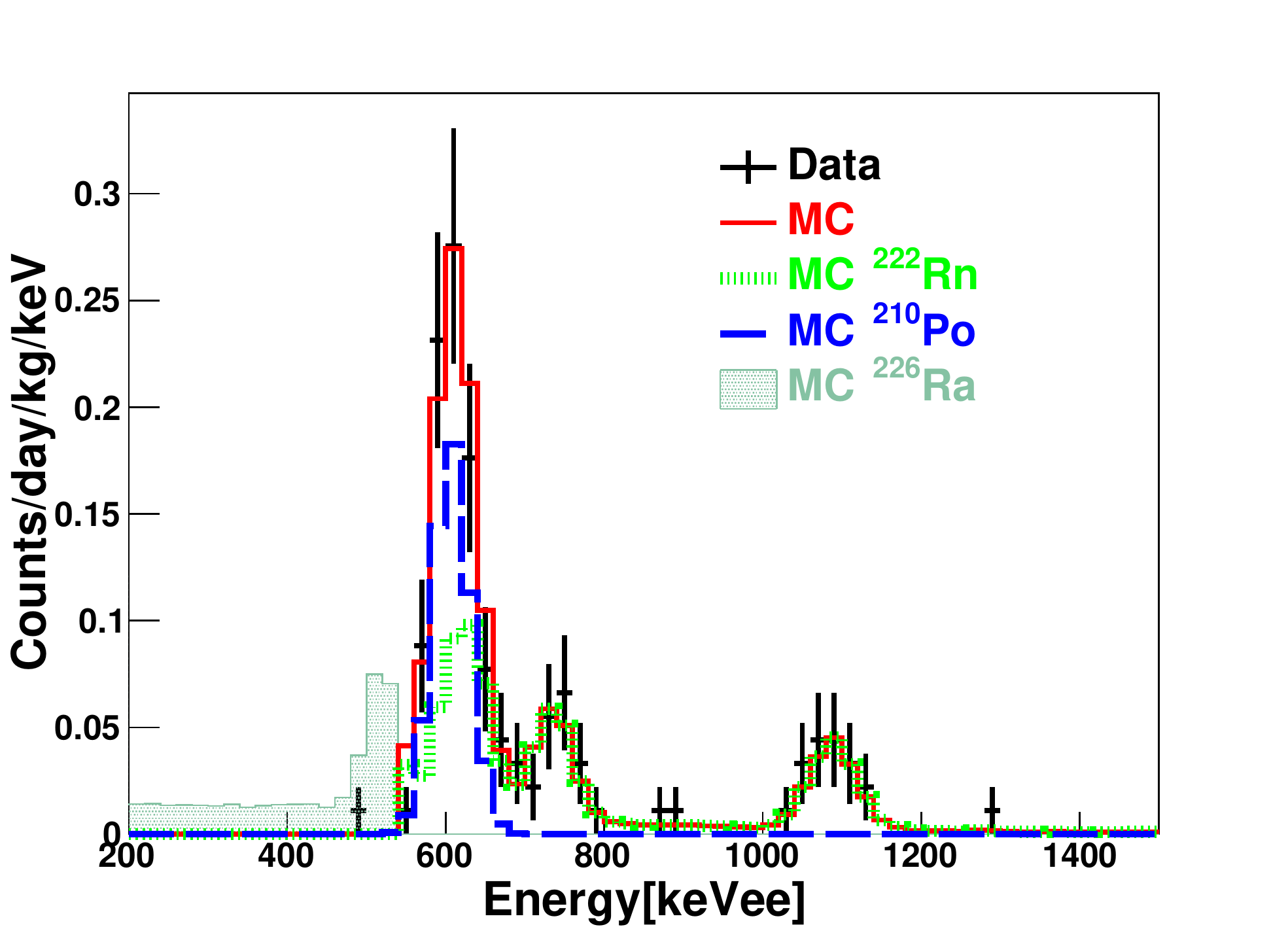}
    \caption{Energy distributions of measured (black) and simulated
      (red) $\alpha$ events. The red line includes the decay chain of
      $^{222}$Rn and $\alpha$ events from $^{210}$Po (blue line). The
      filled area indicates only $\alpha$ events from $^{226}$Ra.}
    \label{fig:sim_dist}
  \end{center}
\end{figure*}

The $\alpha$ energy distribution cannot be fully explained by the
$\alpha$ sources identified by the timing analysis. Since the $\alpha$
source that has $\alpha$ energy slightly less than the 5.59~MeV of
$^{222}$Rn should be added, a decay chain and a $\alpha$-decay are
simulated. The first one is the decay chain beginning with $^{226}$Ra
that decays into 4.87-MeV $\alpha$ and $^{222}$Rn, and the other is
$^{210}$Po, which decays into $^{206}$Pb and 5.41-MeV $\alpha$. As
shown in Fig.\ref{fig:sim_dist}, the remaining component can be
assumed to be $^{210}$Po, and there are no $\alpha$ events from
$^{226}$Ra. This means the possibility of UG-F contamination by the
$^{222}$Rn, so we take data for 13 days, considering the halflife of
$^{222}$Rn. However, the total $\alpha$ rate is stable during the
measurements; therefore,  the LS is contaminated by the
$^{210}$Po. The activities of all components are presented in the next
section, including a comparison with those of the purified sample.

\subsection{Purification of the Liquid Scintillator}
\label{sec:purification}

\begin{figure*}[!b]
  \begin{center}
    \includegraphics[width=.7\columnwidth]{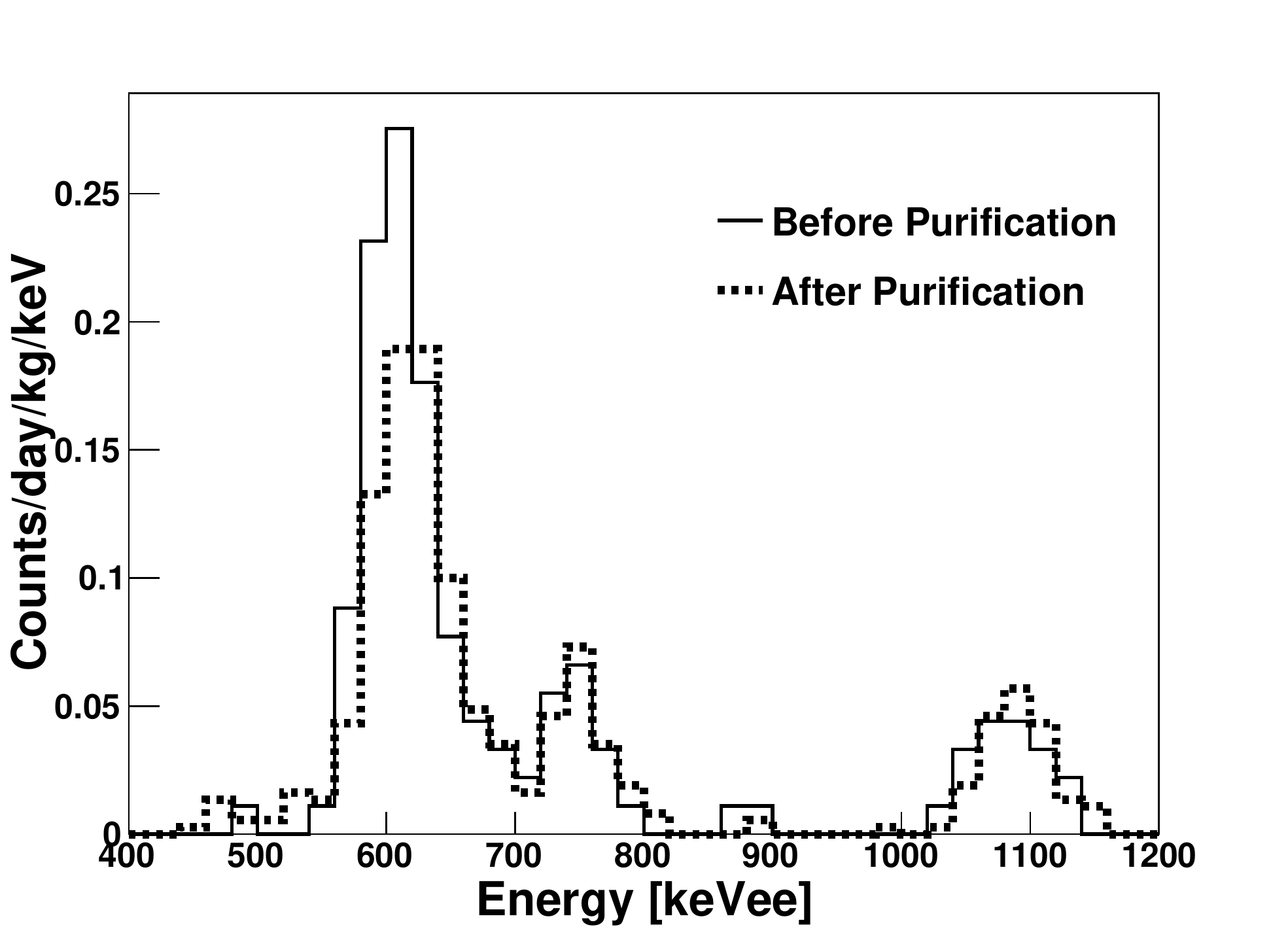}\\ 
    \caption{Comparison of the internal $\alpha$ energy distribution before (black) and after purification (red)}
    \label{fig:alpha_energy}
  \end{center}
\end{figure*}

Purification of the liquid scintillator is a well-known method to
reduce the internal $\alpha$ background of the liquid
scintillator. The combination of alumina adsorption and water
extraction is studied for the UG-F liquid scintillator. The adhesion
of impurities from a liquid to a solid surface is the basic principle
of adsorption. Aluminum oxide, which is an adsorbent, is effective at
removing impurities for metals and ions and has been successfully used
for scintillator purification~\cite{SNO_Quirk}. Alumina powder of
45-$\mu$m grain size and the UG-F are mixed, and are then separated by
vacuum filtration using a PTFE membrane filter with a 0.25-$\mu$m
pore-size. The sample after alumina adsorption is purified once more
by water extraction and nitrogen purging before the measurement. When
two immiscible solvents like water and LS are brought into close phase
contact, impurities present in the LS that are more soluble in water
can be transferred to the water from the LS. Thus, we obtain a
purified LS after re-separating. For the process of water extraction,
the LS and the ultra-pure DI water having a 16.4~M$\Omega$ resistance
are mixed into the container at a ratio of 4:1, respectively. After
mixing, the mixture is stored for a day in order to separate the water
and LS. The LS and water layers are separated using a separation
funnel. Following water extraction, the LS is filled into the
container after nitrogen purging to remove radon contamination during
the purification process.

\begin{table}[t!]
  \centering
  \label{tab:activity}
  \begin{tabular}{c|cccc|c}
    \hline
    Sample & $^{222}$Rn & $^{218}$Po & $^{214}$Po & $^{210}$Po & Total
                                                                 $\alpha$\\
\hline \hline
    ND-1 & 0.030$\pm$0.007 & 0.030$\pm$0.007 & 0.041$\pm$0.008 &
                                                                 0.25$\pm$0.02 & 
                                                                 0.36$\pm$0.04\\
    ND-2 & 0.025$\pm$0.003 & 0.025$\pm$0.003 & 0.032$\pm$0.004 &
                                                                 0.12$\pm$0.007 & 0.21$\pm$0.03\\
    \hline
  \end{tabular}
\caption{Activities of all $\alpha$ components with the UG-F before
  (ND-1) and after (ND-2) purification. All units are mBq/kg.}
\end{table}

Since the LS contamination is estimated to be mostly due to $^{210}$Po
in Sec. \ref{sec:id_alpha}, the LS purification is expected to mainly
reduce $^{210}$Po. We analyze the $\alpha$ events of the sample after
purification via the same method in Sec. \ref{sec:id_alpha}, and the
activities are summarized in Table \ref{tab:activity}. In the table,
ND-1 and ND-2 are the UG-F samples before and after purification,
respectively. As expected, the activity of the $^{210}$Po is reduced
by more than twofold and other components did not show any significant
differences compared to their errors. Fig.\ref{fig:alpha_energy} shows
the energy distributions of the $\alpha$ events before and after
purification.

\subsection{Plan for Main Detector}

\begin{figure*}[!b]
  \begin{center}
    \includegraphics[width=.87\columnwidth]{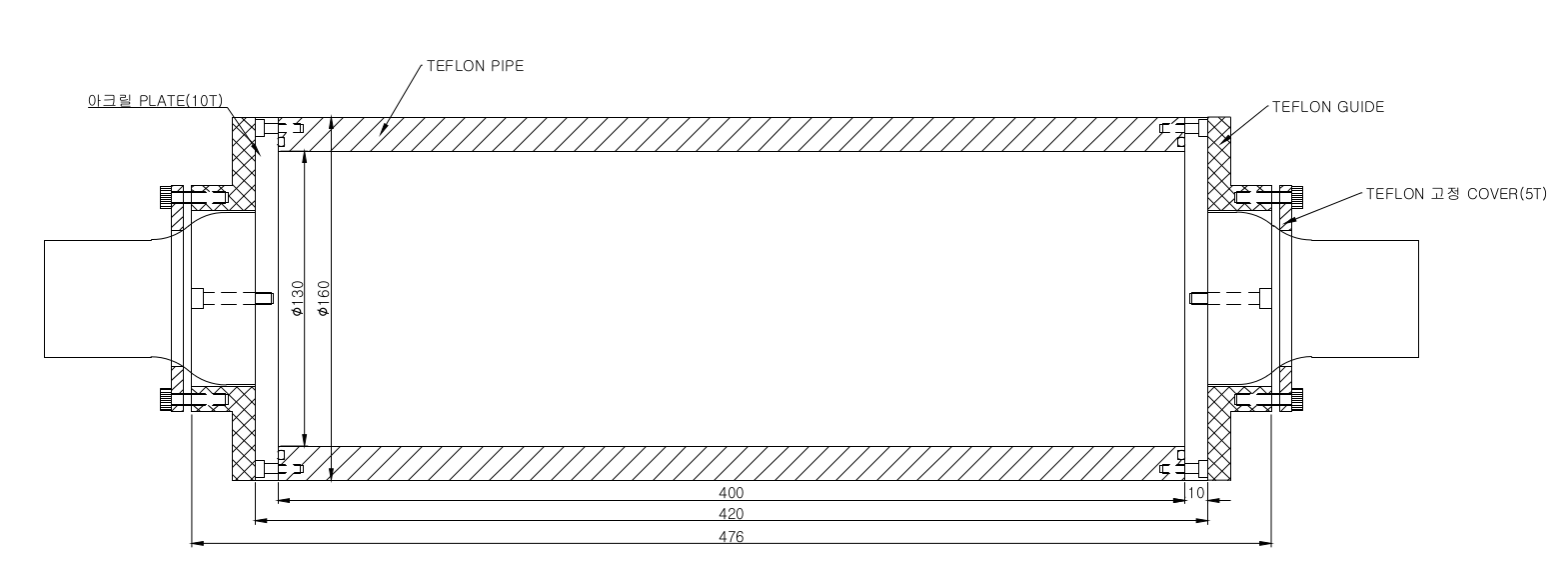}\\ 
    \caption{Design of the COSINE Neutron Monitoring Detector}
    \label{fig:main}
  \end{center}
\end{figure*}

For the main detector, the vessel to be filled with the LS is designed
as a cylindrical homogeneous type made of Teflon with a thickness of
1.5~cm. A homogeneous type has advantages over a segmented type
because spaces and walls between segments may degrade detection
efficiency or lose scintillation photons, which may worsen the
resolution. Due to its higher reflectance, Teflon is considered as the
body of the detector so that we do not need to use any extra materials
to increase light collection efficiency as a reflector. Transparent
acrylic windows with a thickness of 1~cm are coupled to both ends of
the detector with O-rings to prevent the LS from leaking. With limited
available space, the target size is decided to be 40~cm in length and
13~cm in diameter. As a result, the total target volume is about
6~l. Two PMTs are coupled at both ends using optical grease and fixed
using Teflon cover end-caps with a Teflon guide, as shown in
Fig. \ref{fig:main}. The detector is surrounded by dark sheets to
prevent light leaking in. The detector will be installed inside the
main shielding of the COSINE detector.

The detector will be made very soon, and then the detector response,
including PSD properties, will be studied. After understanding the
detector performance, it is going to be installed inside the shielding
of the COSINE detector in the near future. We will also install a
similar type of detector outside of the main shielding to monitor the
fast neutron flux. The detectors will be used to understand the
neutron sources and to study the effects by neutrons in the DAMA
annual modulation study.

\section {Summary}

The LAB- and DIN-based LSs were tested in terms of their PSD
properties to select the target material of the main detector, and the
DIN-based LS, UG-F, is selected due to its better PSD performance
compared to that of the LAB-based LS. We recorded the data with a
prototype neutron detector inside the shielding of the KIMS-CsI at the
Y2L to measure the internal $\alpha$ background. The activity of the
total $\alpha$ events was measured to 0.36$\pm$0.04~mBq/kg, and the LS
contamination is estimated to be almost entirely caused
by$^{210}$Po. The purification process of the LS including water
extraction and alumina adsorption was tested, and the activity of the
$^{210}$Po decreased from 0.25$\pm$0.02~mBq/kg to
0.12$\pm$0.007~mBq/kg.

\acknowledgments
We thank the Korea Hydro and Nuclear Power (KHNP) Company for
providing the underground laboratory space at Yangyang. This work is
supported by the Institute for Basic Science (IBS), Republic of Korea,
under project code IBSR016-A1.

\end{document}